# Organization studies Based Appraisal of Institutional Propositions in the Nigeria Data Protection Regulation

**Sabo, Sumayya Babangida & Utulu, Samuel C. Avemaria**
Department of Information Systems,
School of Information Technology and Computing
American University of Nigeria, Yola, Adamawa State, Nigeria
**E-mail**: Samuel.utulu@aun.edu.ng; sabo.su@aun.edu.ng
**Phones**: +2348163525341; +2348036083887

## ABSTRACT

The central notion in the organization studies is that the formation, sustenance and demise of organizations are directly connected to the socio-cultural contexts where they are situated. So, the extent organizations are able to handle the challenges to personal data management can be linked to the socio-cultural contexts where they are situated. This justifies the reasons why regulations such as the Nigeria Data Protection Regulation (NDPR) are produced to initiate institutional propositions that shape socio-cultural contexts of organizations into forms that promote efficient personal data management. This study, a part of a larger study, appraises the NDPR in order to assess how the institutional propositions in the regulation can help shape the socio-cultural contexts of organizations in Nigeria, and then, promote personal data management practices. The study uses the narrative literature method to select the publications that were used to reach its objectives. The study addresses the dearth of scientific research that describes how institutional propositions in the NDPR can impact on the socio-cultural contexts of organizations in Nigeria. It also ameliorates gaps in scientific knowledge about how institutional propositions in the NDPR can be legitimized and help restructure socio-cultural contexts of organizations in Nigeria, and in effect, promote efficient personal data management practices.

1.. INTRODUCTION

Many organizations including, governments, intergovernmental organizations, non-governmental organizations (NGOs), and for-profit organizations are cashing in on the value of data to gain strategic advantage. Organizations have evolved from the traditional database management systems era to a more sophisticated data analytics era. In the data analytics era, organizations collect, store, process and use data for strategic business decision making (Alsghaier et al., 2017). New sophisticated hardware, software and algorithms enable organizations to use machine learning (ML) and artificial intelligent (AI) tools to analyze vast number of data. ML and AI enable contemporary organizations to produce reliable predictive and prescriptive models that proffer solutions to complex strategic business and social challenges (Alsghaier et al. 2017; Lopenioti et al. 2020; Sutduean et al., 2019). The implication of evolving forms of organizing is that peoples' personal data are collected and used across vast online business management platforms for diverse purposes. Interestingly, the growth in the collection and use of people's personal data is not limited to for-profit business organizations. The development has also been traced to governments, their agencies and NGOs and indicate that they are the highest collectors and users of peoples' personal data (Gazi, 2020). This is the reason why it has become mandatory for governments, intergovernmental organizations, NGOs, and for-profit organizations, to have well laid out personal data management plans.

Contemporary realities show that everyone is prone to the menace caused by the limitations in organizations' ability to handle data management challenges (Gazi, 2020). The role of socio-cultural contexts of organizations in organizations' ability to handle personal data management challenges is imperative. Socio-cultural contexts of organizations comprise of legitimized rules, regulations and cultures that legitimize organizational practice and practices (Svenson & Freiling, 2019; Greenwood et al., 2017; Meyer & Rowan, 1977; Scott, 1995). Legitimized practice and practices defines power relations, internal functioning and the relationships among organizations (Greenwood et al., 2017). Consequently, socio-cultural contexts of organizations come to bear in the ability of organizations to handle data management challenges (Kasim et al., 2022; Utulu & Ngwenyama, 2021). The notion makes governments, inter-governmental organizations and other stakeholders across the globe to put forward personal data management laws and regulations to guide the structuring of socio-cultural contexts of organizations to facilitate efficient personal data management.

In Nigeria for instance, there are a number of laws and regulations that were produced to help structure socio-cultural contexts of organizations. Accordingly, the NDPR was produced based on "…concerns and contributions of stakeholders on the issues of privacy and protection of Personal Data and the grave consequences of leaving Personal Data processing unregulated (NITDA, 2019, p. 3)." Consequently, the National Information Technology Development Agency Act of 2007, Cyber security Act of 2015 and NDPR, 2019 were produced to help restructure the socio-cultural context of organizations in Nigeria (FGN, 2007; FGN, 2015; NITDA, 2019). However, the lack of understanding of how to successfully implement the laws and regulations still constitutes critical challenges in Nigeria. We ascribe this persisting limitation in the implementation of the NDPR to the dearth of scholarly research into the relationships among the NDPR, organizations and the socio-cultural contexts of organizations in Nigeria.

Consequently, this study appraises the NDPR using perspectives in the organization studies field as it theoretical lens and is informed by the following questions: what are the attributes of the institutional propositions in the NDPR and how can they be legitimized to restructure the socio-cultural contexts of organizations in Nigeria, and in effect promote efficient personal data management?

### 1.1. Nigeria Data Protection Regulation

The NDPR was produced in 2019 by the National Information Technology Development Agency (NITDA) as part of NITDA's statutory responsibility toward productive use of IT and personal data protection in Nigeria (Greenleaf, 2019; NITDA (2019). The objectives for producing the NDPR are, to safeguard rig hts of data subjects to data security and privacy, inject safety measures to transactions involving exchange of personal data, avert personal data manipulation, and ensure continuous and inclusion of Nigerian organizations among those that implement personal data management best practices (NITDA, 2019, p.4). " The regulation is made up of four parts namely, preamble, operational propositions, rights of data subjects, and institutional propositions. The primary motive behind the production of the NDPR is to regulate organizations. Invariably, regulating organizations has to do with initiating organizational rules of engagement that bring to bear new institutions that produce new organizational forms (Deephouse & Suchman, 2008; Greenwood et al., 2017; Scott, 1995).

For instance, the operational propositions initiated in the NDPR promote five operational processes required to handle personal data management (See Babandiga Sabo et al. 2023). The five operational processes include, building organizational practices that encompass coming up with data protection government principles. Second is setting up prerequisite practices including, initiating with publicity and privacy policy for personal data management, declaration of motives, and initiating reliable data security measures. Third is initiating ethically and legally acceptable personal data collection methods which ensures that data subjects' informed consent is sought for and that personal data are collected specifically for the purposes consented for. Fourth is initiating ethically and legally acceptable personal data processing and use practices including, spelling out how third party comes into bear in personal data processing and use. Fifth is spelling out liabilities due organizations (data controllers) and the benefits they stand to gain from appropriate personal data management.

### 2. LITERATURE REVIEW

Organizations have been the bedrock of human organizing from antiquity to the present age. Organizations have been defined as collections of people who are working towards achieving the same purpose (Daft, 2015). In trying to describe the complexities surrounding legitimizing organizational practice and practices, Daft, (2015) posits that the "…specific challenges today's…organizations face are globalization, intense competition, rigorous ethical scrutiny, the need for rapid response, the digital workplace, and increasing diversity (p. 7)." In trying to establish how organizations are legitimized, scholars in the organization studies field address three broad questions: what propels the formation of organizations, sustenance and demise. To address these questions, three theoretical strands namely, the ecological, institutional and interpretivist strands were developed.

The three strands provide different ways for viewing the nature of organizations and hence, different, but somewhat similar views, about what constitutes the right answers to formation, sustenance and demise of organizations. The ecology strand adopts insights in biology, sociology, economics and statistics to analyze organizational diversity at three levels namely, community, population and organization levels. Singh & Lumsden, (1990) argue that the ecology strand's "…key concerns are to investigate how social conditions influence (a) the rates of creation of new organizational forms and new organizations, (b) the rates of demise of organizational forms and organizations, and (c) the rates of change in organizational forms (p. 162)."

Organizational ecologists will have us believe that organizations are a conglomeration of organizations formed into a community and population that constitute social systems called organizational environment. Baum & Amburgey (2017) therefore opine that the strand "…aims to explain how social, economic and political conditions affect the relative abundance and diversity of organizations and to account for their changing composition over time (p. 304)." The terms, social, economic and political conditions as used in Baum & Amburgey (2017) are broader categorization of the social conditions listed in Daft, (2015): globalization, intense competition, rigorous ethical scrutiny, the need for rapid response, the digital workplace, and increasing diversity are all forms of social, economic and political conditions.

The production of NDPR in 2019 can be directly ascribed to these three social conditions given NITDA's (2019) testimony that the NDPR was produced taking "Cognizant of emerging data protection regulations within the international community… (p. 3)." Conditions like this make organizational ecologists to argue that the quality of interactions and relationship an organization have with other organizations within the social system where it operates determines its sustenance and demise (Baum & Amburgey, 2017; Todd et al., 2014). Unfortunately, there is no study, as far as we can confirm, that aims at using theoretical insights in organizational ecology to evaluate how to drive the implementation of the NDPR among organizations. The second theoretical strand in the organization studies is the interpretivist strand and it pays attention to internal organizational environment and sees it as socially constructed by organizational actors (Todd et al., 2014; Dandridge et al., 1980; Svenson & Freiling, 2019).

Invariably, scholars following the organizational interpretivism approach see organization as culture and they promote the notion that the cultures are socially constructed and contained in people's cognition (Daniels & Johnson, 2002; Tyler & Gnyawali, 2009). Organizational interpretivism pay credence to histories of organizational realities (Rowlinson & Procter, 1999; Suddaby & Foster, 2017; Utulu, 2019; Utulu & Ngwenyama, 2021). Scholars who study organizations from the perspective of organizational interpretivism also believe that organizations are immersed into the social contexts where they are located. And by this, they do historical evaluation on how organizations' interactions with the social contexts impact on their international functioning and existence (El Sawy et al., 1986; Rowlinson, 2020; Utulu, 2019). It is unfortunate, therefore, that scholars have not produced research works that depict how cultural traits in the Nigerian organizational environments impacts on the adoption of the NDPR.

Organizational institutionalism is the third strand and uses theoretical perspectives in the institutional theory to explain how organizations emerge, sustained and demise. It uses the theory to evaluate how socially constructed institutions legitimize actions taken by organizations, and how the institutions evolve, exist, and get discarded over time (Boxenbaum, 2014; Greenwood et al., 2017; Lok et al., 2017). In the institutional theory, institutions are taken to be rules, norms, regulations, etc. They serve as the threads that hold social structures together and provide meanings used to determine why social behaviors are acceptable or not. Meyer and Rowan (1977) proposes how realities within organizations are socially constructed by a multiple of social actors including those within and outside organizations. Scott (1995), asserts that "[i]nstitutions…are composed of cultural-cognitive, normative, and regulative elements that, together with associated activities and resources, provide stability and meaning to social life…[they] are transmitted by various types of carriers, including symbolic systems, relational systems, routines, and facts (p. 33)."

Given the believe that realities within organizations are not only triggered by social actors within them but also by social actors outside them, the organizational institutionalism strand therefore shows interest in both internal and external organizational environments. While they study institutions within organizations, they also pay attention to how institutions outside organizations influence organizational forms (Deephouse & Suchman, 2008; Greenwood et al., 2017; S. C. Utulu, 2019). Going by the theoretical notions in the organization studies field, it can be deduced that the successfully implementation of the NDPR among organizations in Nigeria is not just the problem of each organization, but also that of the organizations that populates its external environment.

## 3. RESEARCH METHODOLOGY

The aim of the study is to appraise the NDPR and assess how it can help shape the socio-cultural contexts of organizations in Nigeria, and in effect, promote personal data management practices. To achieve its aim, the study uses the theoretical insights in the institutional theory to interpret the NDPR. To achieve the objective and aim, two subjective literature review methods were adopted namely, the narrative literature review and snowball literature review. The narrative literature review is one of the scientific literature review methods available to scholars and it is distinctive from other literature review methods given its subjective approach (Utulu et al. 2013; Avgerou, 2010; Baumeister & Leary, 1997).

We adopted Avgerou's (2008) narrative literature review framework which comprises of four steps namely, having "an 'insider's' confidence of understanding the unfolding of this research area and its literature (p. 134)". Second, make sense of the literature and substantiate the validity of arguments and descriptions with references. Third, avoid vulgar eclecticism and 'inbreeding'. Fourth, do exhaustive reading and reading of selected literature. This study adopts these steps and benefited from the authors' experiences of dealing with data subjects and personal data. We made reference to literature in the course of study and validated our arguments and descriptions with references mainly retrieved from Google Scholar.

## 4. SOCIO-CULTURAL CONTEXTS OF ORGANIZATIONS IN NIGERIA AND THE NIGERIA DATA PROTECTION REGULATION

The notion propagated in this study is that socio-cultural contexts of organizations are crucial to the implementation of personal data management among organizations. And that theoretical insights in the organization studies field are sine-qua-non to understanding the connections among institutional propositions in the NDPR, socio-cultural contexts of organizations in Nigeria and the implementation of the NDPR among organizations in Nigeria (Baum & Amburgey, 2017; R. Todd et al., 2014; Singh & Lumsden, 1990b; Svenson & Freiling, 2019). It is argued that the three theoretical stances dominant in the organization studies field are relevant to ongoing debate on the implementation of the NDPR. This is because data protection regulation issues revolve around organizations that are situated in socio-cultural organizational environments. The organizations include, intergovernmental organizations, governmental organizations, NGOs and privately owned organizations.

Invariably, the value of interactions among organizations within any socio-cultural organizational contexts primarily determine the extent they meet their goals (Greenwood et al., 2017; S. C. Utulu, 2019). As shown in Figure 1, issues surrounding the implementation of NDPR in Nigeria can be traced from global organizational context through regional organizational context and national organizational context to specific organizational contexts. This is consistent with positions postulated in Giddens, 1991) and aligns with the historical analysis organizational interpretivism scholars engage in with the aim of revealing how organizational external contexts impact organizational internal contexts (Dandridge et al., 1980; Svenson & Freiling, 2019).

Insights represented in Figure 1 is also consistent with positions promoted in the organizational ecology strand. Aside looking at the entire organizational socio-cultural contexts which include both global and regional contexts, Figure 1 shows that the national contexts provides grounds for interrelationship among multiple organizations. The organizational ecology strand views organizations the same ways biologists view biological ecology in order to expose and evaluate how the interrelationships among organizations define organizational ecology (Baum & Amburgey, 2017; Singh & Lumsden, 1990).

Figure 1 shows that the national socio-cultural organizational context comprise of the national data protection regulatory legal frameworks. Public and private organizations including, the legislative, judiciary and privately owned legal firms provide leadership when it comes to ensuring that laws and legislations are used to drive how organizations handle personal data protection (Greenleaf, 2019; Salami, 2020). (There is also the declaration in the NDPR (2019) that organizations "… shall ensure continuous capacity building for Data Protection Officers and the generality of her personnel involved in any form of data processing (p. 18)." Aside this, the regulation also talked about the registration and licensing of Data Complaint Organizations that shall "…audit, conduct training and data protection compliance consulting to all [organizations] (NITDA, 2019, p. 18)."

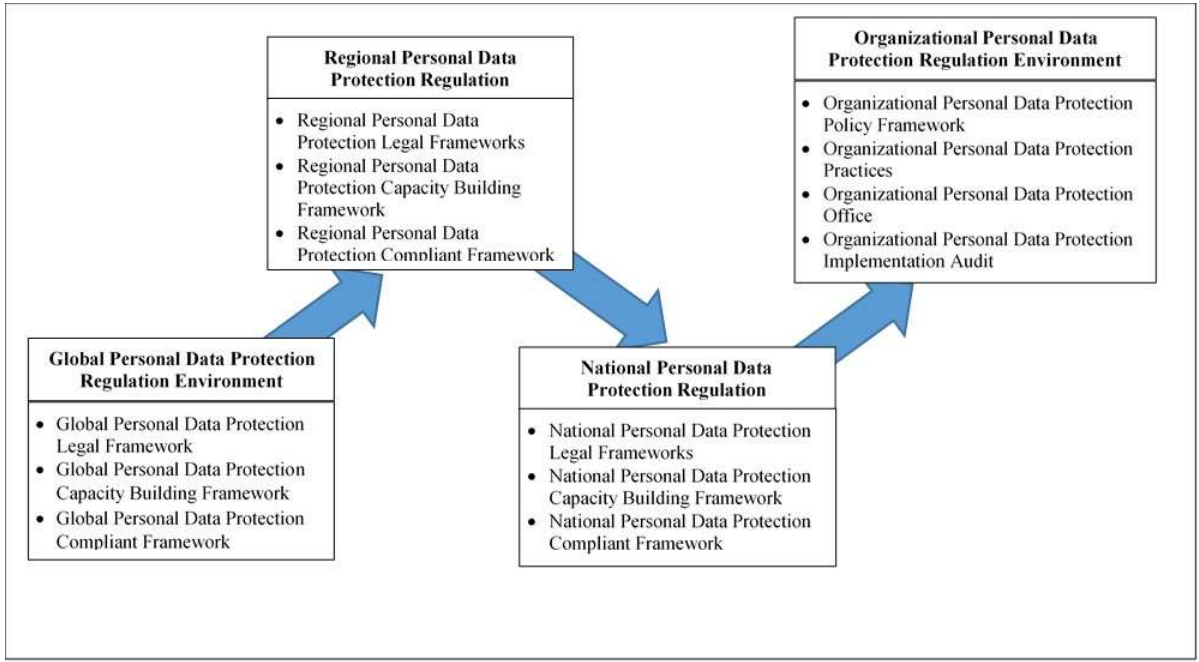

**Figure 1: Dynamics of Institutional Arrangements for Personal Data Protection Regulation**

The implication of this is that the national personal data protection regulation context comprises of numerous organizations serving different functions. The scenario also indicates the relevance of the organizational institutionalism strand. The organizational institutionalism devotes its evaluation to exposing and assessing how institutions that legitimize organizational actions are produced, sustained and discarded (Meyer & Rowan, 1977; Rowlinson & Procter, 1999; Scott, 1995; Suddaby & Foster, 2017). The organization level personal data regulation context depicted in Figure 1 succinctly aligns with notions propagated in both the organizational interpretivism and organizational institutionalism strands. Aside showing how the NDPR provides grounds for restructuring internal structures of organizations by introducing the office of Personal Data Protection Officer, it also reveals that the NDPR requires the creation of new organizations namely, complaint organizations.

This is a typical example of the emergence of new organization forms and new organizations as depicted in the organization studies field (Baum & Amburgey, 2017; Greenwood et al., 2017; Rowlinson & Procter, 1999; Scott, 1995; Singh & Lumsden, 1990b). Organizational symbolism is mirrored in that part of Figure 1 and provides grounds for evaluating the emergence and institutionalization of new organizational practices. Organizational symbolism, myth and institutionalization process all form core aspects of organizational interpretivism and organizational institutionalism (Berger & Luckmann, 1967; Deephouse & Suchman, 2008; Meyer & Rowan, 1977; Utulu & Ngwenyama, 2021). So, research questions relating to the extent the NDPR is designed to reconstitute the socio-cultural contexts of organizations in Nigeria is very relevant to ongoing efforts to efficiently and effectively implement the NDPR.

## 5. CONCLUSION AND LIMITATIONS OF THE STUDY

The aim of the study is to appraise the NDPR and assess how it can help shape the socio-cultural contexts of organizations in Nigeria, and in effect, promote personal data management practices. The need for the study arises given that many of the studies that have been published on the NDPR do not address its value to personal data management from perspectives inherent in the organization studies field. In other words, scholars do not see issues causing the low value accorded to the NDPR in Nigeria among organizations from organization studies perspectives. Consequently, this study presents the relevance of perspectives in the organization studies field to assessing and understanding the value of the NDPR.

It is hoped that the study will present the knowledge required for growing the implementation of the NDPR among organizations in Nigeria. The study concludes that issues affecting the implementation of the NDPR among organizations are organization oriented issues and consequently, should be addressed as such. Like every other studies, the study has some limitations. The foremost limitation of the study is that it is totally based on narrative and snowball literature review methods. In other words, the study did not use empirically derived data to explain the positions that were put forward in it. It is imperative to however, state that this limitation did not rob the study of its theoretical and practical values.

Proceedings of the Cyber Secure Nigeria Conference – 2023